\documentclass{blois}

\def\be{\begin{equation}}
\def\ee{\end{equation}}
\def\bea{\begin{eqnarray}}
\def\eea{\end{eqnarray}}

\begin{document}

\begin{center}

\vspace{2cm}

{\Large \bf LIGHT STOPS FROM EXTRA DIMENSIONS}

\vspace{0.5cm}

MATEO GARC\'IA PEPIN 

\vspace{0.2cm}
{\em Institut de F\'isica d'Altes Energies (IFAE),
The Barcelona Institute of Science and Technology (BIST),
Campus UAB, 08193 Bellaterra (Barcelona) Spain.}
\end{center}
%\maketitle
\vspace{0.5cm}
\abstracts{
In supersymmetric models the mass of the stops can be considered as the naturalness measure of the
theory. Roughly, the lighter the stops are, the more natural the theory is. Both, the absence of supersymmetric
signals at experiment and the measurement of the Higgs mass, put scenarios with light stops under
increasing tension. I will present a supersymmetry breaking mechanism of the Scherk-Schwarz
type that, by introducing extra $SU(2)_L$ triplets in the Higgs sector, is able to generate the correct Higgs
mass while keeping stops light.}

\section{Introduction}
In BSM theories and because of the strength of the top yukawa coupling, top partners
can be (roughly) considered as a measure of the naturalness of the theory. In the particular case of supersymmetry these top partners are the scalar top quarks, which, when appearing, are expected to cancel the dangerous contributions to the Higgs potential that destabilize the Electroweak VEV. A natural version of supersymmetry as a UV completion of the SM would therefore require stops to be as light as possible to minimize fine-tuning.

However, light stops (below $1$ TeV) are difficult to encompass within the current experimental situation. Through dominant 1-loop contributions, stop masses and the mixing in the stop sector are critical in the predicted value of the Higgs mass. They are required to be as large as possible in order to fit the $125$ GeV experimental measurement~\cite{Delgado:2013gza}. On top of that, searches for superpartners at the LHC have returned null results up to now. Since common models of soft SUSY breaking and RGE correlations tend to relate the masses of superpartners, bounds on other sparticles will constrain the SUSY breaking
mechanism, affecting the value of the stop mass. For instance, normally, the heavier the gluino, the heavier the stop.

As the quest for a natural theory of the EW scale by means of supersymmetry is becoming harder, it is mandatory for BSM model builders to explore non minimal realizations of supersymmetry. This was our aim when working on the model that I now present. It was introduced in Ref.~\cite{Delgado:2016vib}, to which, together with Refs.~\cite{Pomarol:1998sd,Delgado:1998qr}, I refer the reader for technical details.

\section{The model}
We start by embedding the MSSM in a 5D space-time setup where the extra dimension
 is the orbifold $S^1/Z _2$ with two four-dimensional (4D)
 branes at the fixed points $y=0$ and $y=\pi R$~\footnote{$R$ is the radius of
 the circle $S^1$.}.  The gauge and Higgs sectors, as well as the first
 and second generation of matter (and the right-handed
 stau~\footnote{We are considering $\widetilde\tau_R$ propagating in
   the bulk in order to avoid bounds on heavy stable charged
   particles~\cite{CMS:2016ybj}.
   %, as we will comment later on.
 }),
 propagate in the bulk while the rest of the third generation matter
 is localized at the $y=0$ brane. A Scherk-Schwarz twist breaks N=1 Supersymmetry. It consists in imposing to the (superpartner) 
5D fields a nontrivial twist ($\omega$) under a $2\pi$ translation on the fifth dimension. This breaking felt by the bulk fields will then be translated to the fields on the brane radiatively. These are the main features of the scenario:
\begin{itemize} 
   \item
     The Higgsino zero mode has a Dirac mass equal to $\omega/R$, by
     which there is no need of introducing a superpotential $\mu$-like
     term as in the MSSM. There is no $\mu$-problem.
  \item
     The lightest $(n=0)$ modes of the fields in the bulk have tree-level
     masses that are either zero, $\omega/R$, $2\omega/R$ or
     $1/R$ (see Fig.~\ref{fig:lowen}). Vanishing masses correspond to SM-like fields.
     \begin{figure}
\begin{minipage}{0.5\linewidth}
\centerline{\includegraphics[width=0.8\linewidth]{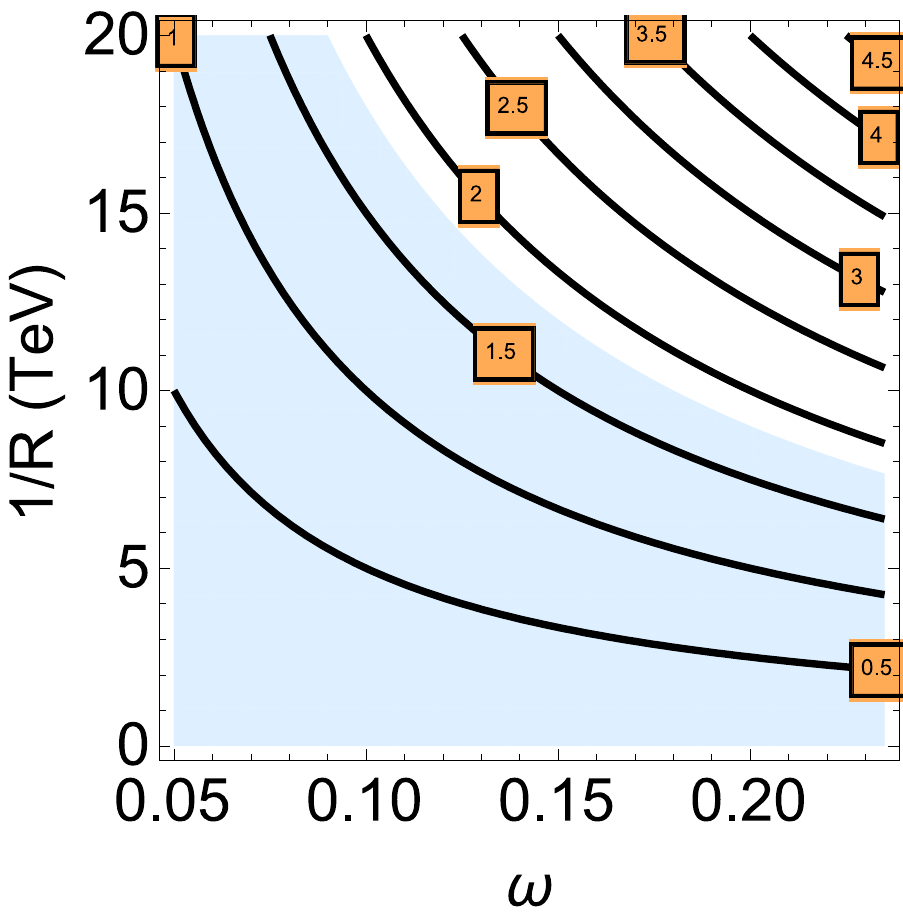}}
\end{minipage}
\hfill
\begin{minipage}{0.5\linewidth}
\centerline{\includegraphics[width=0.8\linewidth]{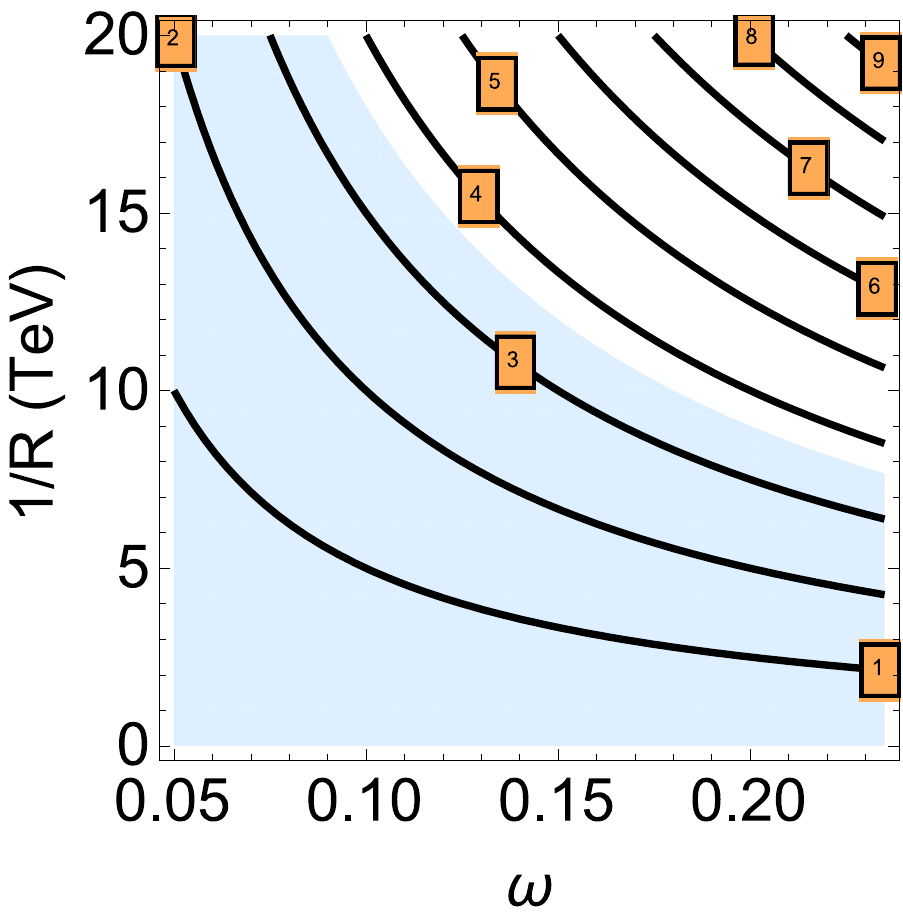}}
\end{minipage}
\caption[]{Contour plots of the tree-level masses of the bulk
    fields sensitive to the SS mechanism. Labels are in TeV units.
    Left panel: First and second generation sfermions, right-handed
    stau, gauginos and Higgsinos. Right panel: Charged and neutral
    heavy doublet Higgses. In light blue the region with gluino mass
    $m_{\tilde{g}}<1.8$ TeV, in tension with LHC bounds.}
\label{fig:lowen}
\end{figure}
  \item
     States localized in the brane, are naturally light as
     their tree-level masses are vanishing.  Their one-loop radiative
     masses from KK modes are finite and can be
     interpreted as finite threshold effects after integrating out the
     heavy modes (see left panel of Fig.~\ref{fig:stops}). 
     \begin{figure}
\begin{minipage}{0.5\linewidth}
\centerline{\includegraphics[width=0.8\linewidth]{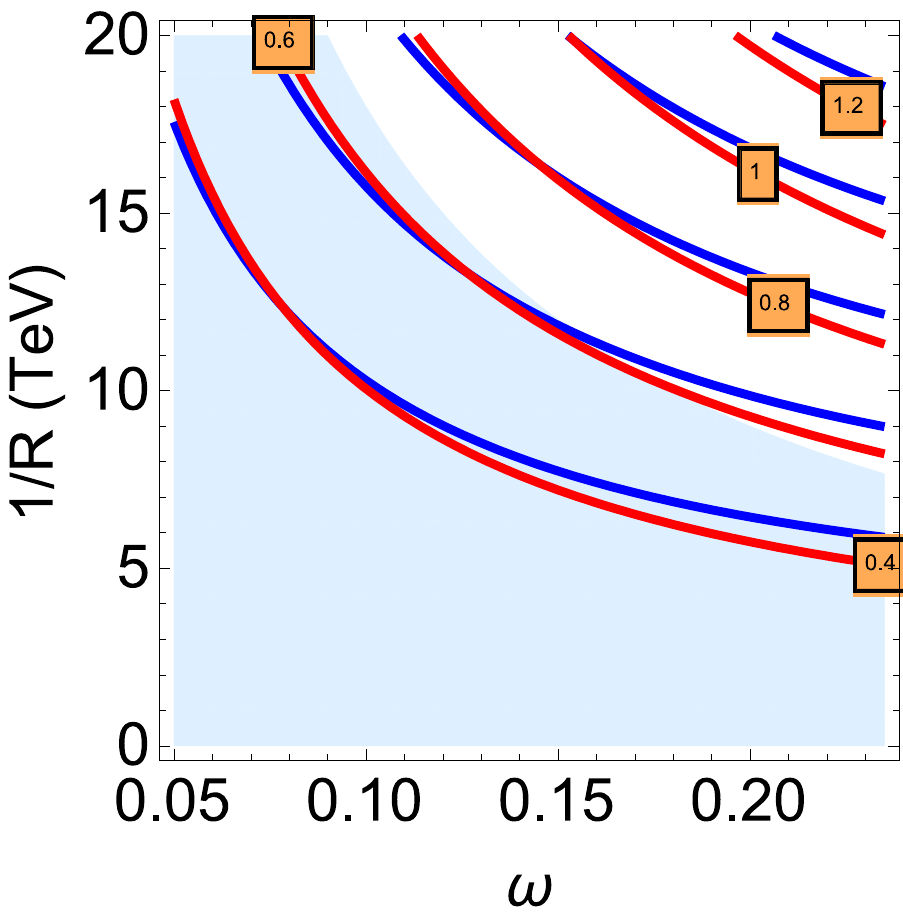}}
\end{minipage}
\hfill
\begin{minipage}{0.5\linewidth}
\centerline{\includegraphics[width=0.8\linewidth]{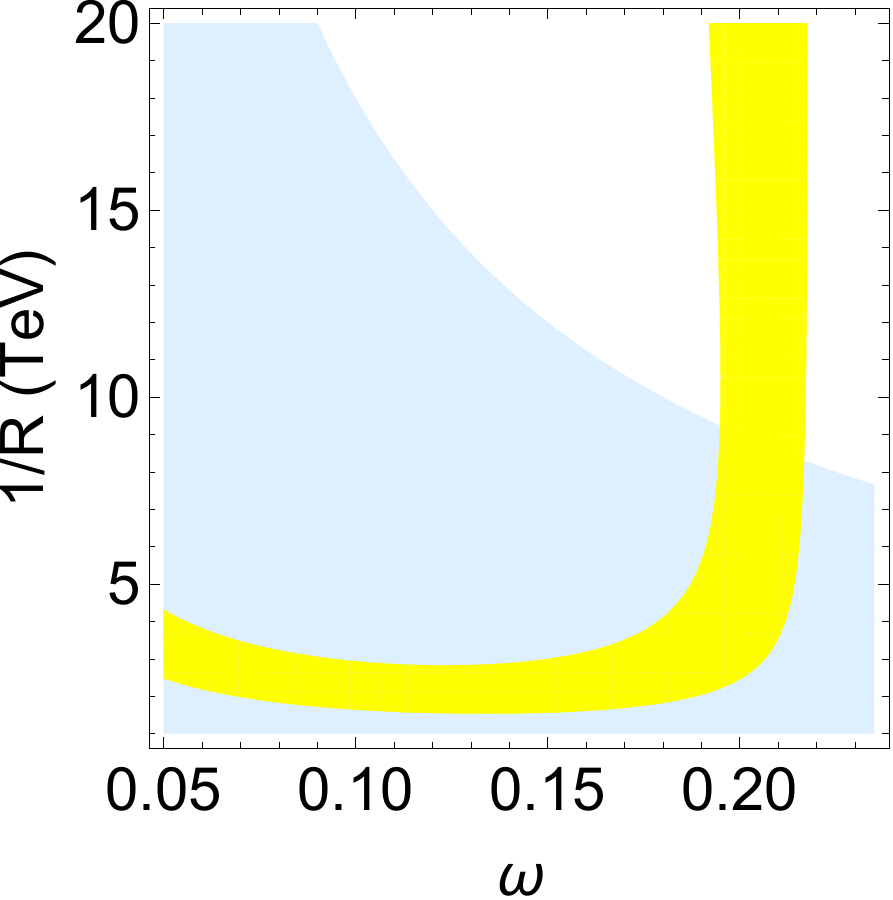}}
\end{minipage}
\caption[]{ Left panel: Masses of the lightest stop $\tilde{t}_1$
and sbottom $\tilde{b}_1$ (red and blue lines respectively). Right panel: The parameter space
of the $(\omega, 1/R)$ plane (yellow area) where the experimental EWSB with the correct Higgs
mass is successfully achieved. In light blue the region with gluino mass
    $m_{\tilde{g}}<1.8$ TeV, in tension with LHC bounds.}
\label{fig:stops}
\end{figure}
   \item
     At tree-level the theory predicts a 4D massless Higgs doublet
     with a flat potential while the rest of the
     Higgs sector is heavy. 
\end{itemize}

The last point is what generates the drawbacks of the minimal picture in the SS SUSY breaking paradigm. The 4D effective potential that is left when integrating out KK modes is the following, 
\be
V_{SM}= (m_{\mathrm{tree}}^2+\Delta_h m^2) |h^{(0)}|^2+\left(\lambda_{\mathrm{tree}}+\Delta\lambda\right)|h^{(0)}|^4+\dots \, ,
\label{eq:Vtrip}
\ee
where the value $m_{\mathrm{tree}}$ depends on the choice of SS twist (in our case $m_{\mathrm{tree}}=0$) and $\lambda_{\mathrm{tree}}$ on the structure of the Higgs sector. $\Delta_h m^2$ and $\Delta \lambda$ correspond to loop order contributions. 

From the potential shown above we can identify two issues:
 \begin{description}
 \item[i) EWSB:] As stated above, since we have $m_{\mathrm{tree}}=0$, EWSB has to proceed via radiative corrections. At one-loop, there are
   gauge corrections that feed $\Delta_h m^2$ which are positive, thus preventing
   EWSB~\footnote{Also, stops are localized, thus massless at tree level and they do
   not produce any 1-loop correction to the Higgs mass
   proportional to $h_t^2$ which could trigger EWSB as in the 4D
   MSSM.}. 
\item[ii) Higgs mass:] As both stop soft masses and trilinear stop mixing parameter
  are 1-loop suppressed, their radiative correction to the Higgs
  quartic coupling is too small to reproduce the experimental value
  $m_h\simeq 125\,$GeV.  
 \end{description}

To solve the EWSB and Higgs problems we consider the scenario where the Higgs sector is extended by
hyperchargeless $SU(2)_L$ triplets propagating in the bulk~\footnote{As we refrain from introducing any dimensionful
parameter in the 4D superpotential, we do not consider the option of
triplets localized on the brane. In which case the fermionic triplet components
would be too light to overcome the chargino mass bound~\cite{Heister:2002mn}.}. It turns out that triplets introduce new radiative corrections to the mass of the Higgs doublet which are negative and help in ending up with a tachyonic mass for the Higgs doublet, thus facilitating the observed EWSB in a somewhat wide parameter region (see right panel of Fig.~\ref{fig:stops}). Also, the presence of triplets enhances the tree-level Higgs quartic coupling in the effective theory
and thus makes it easy to accommodate the 125-GeV Higgs mass
constraint. 

\section{The low energy theory}
Below the energy scale of the bulk fields with masses $\mathcal{O}(\omega/R)$ we are left with
\be
\mathrm{SM} + \sigma^{(0)} + \tilde{\ell}_3 + \tilde{t}_{1,2} + \tilde{b}_{1,2}\, .
\ee
In this setup the tau sneutrino ($\widetilde\nu_\tau$) is the LSP. The $\widetilde\nu_\tau$ is not a
good dark matter candidate as it would provide the observed relic abundance
for a (small) mass range that is nevertheless ruled out by direct detection
experiments~\cite{Falk:1994es}. In the remaining mass region, its relic
density has to be somehow reduced~\footnote{This is possible if the sneutrinos
do not reach thermal equilibrium before their freeze out, or an
entropy injection occurs at late times (see
e.g. Ref.~\cite{Gelmini:2010zh}). Alternatively, decays such as $\widetilde
\nu_{\tau}\rightarrow \tau\bar{\tau}$
can provide the desired dilution. These could in principle be
generated by operators like $LLE$ that introduce a small R-parity
violation.}.

Even though the gluino is not part of the low-energy theory, the most
robust constraint to the parameter space of the model is provided by
the gluino direct searches. From early 13-TeV data, ATLAS and CMS set
the bound on $m_{\tilde{g}}$ at $1.8$ TeV~\cite{ATLAS-CONF-2015-067,CMS-PAS-SUS-15-003}. Since the
whole spectrum mostly depends on just two parameters, $\omega$ and
$1/R$, and in particular $m_{\tilde{g}}=\omega/R$, the gluino mass
bound constrains the low energy theory. In particular, the
gluino bound forces the mass of the stops and sbottoms to be roughly
above $550$\,GeV and the scalar triplet, stau and tau sneutrino to be
heavier than around $250$\,GeV. Of course, by excluding heavier gluino
masses we will also be able to set stronger bounds on third-generation
squarks, the triplet and the stau doublet. 

As we are dealing with a heavy LSP with a mass typically above
$300$\,GeV, the LHC bounds from stop searches are very mild or even
absent~\cite{Aad:2015pfx}. In addition, considering usual bounds is a
conservative assumption; in this model the topology of the stop decays
is different from what is expected in MSSM-like scenarios. Because the
stop is lighter than all neutralinos and charginos, it decays to
off-shell states such that the final signature is a multi-body decay
for which the current stop bounds can be very much
softened~\cite{Alves:2013wra}. Bounds on sbottoms are more severe than
those on stops (for LSP masses below $400$ GeV, ATLAS and CMS exclude
sbottom masses up to
$900$\,GeV~\cite{ATLAS-CONF-2015-066,CMS-PAS-SUS-16-001}) but they
suffer from the same softening mentioned above for stops

\section{Summary}
I have shown how extra dimensions, by means of Scherk-Schwarz SUSY breaking, can serve as a tool to 
minimize the fine-tuning triggered by the LHC constraints on minimal
supersymmetric extensions of the Standard Model. In order to solve the EWSB and Higgs mass drawbacks of the Scherk-Schwarz paradigm, $Y=0$ $SU(2)_L$ triplets propagating in the bulk are introduced, these provide radiative corrections that trigger EWSB and enhance the tree level Higgs mass.

Due to the mass hierarchy between fields that propagate in the bulk and fields localized in the brane, most of the new-physics sector is decoupled
from EW-scale processes, in agreement with
experiments. However some superpartners, tightly linked to naturalness
and/or properties of the Scherk-Schwarz twists, have to be light and
populate the low-energy particle content of the theory, which
eventually consists of the Standard Model degrees of freedom plus a
scalar triplet, the third-generation of squarks and the doublet of sleptons. 

Since gluino bounds are robust and quite generic, the most stringent
constraint to the model comes from gluino searches. Nevertheless,
other experimental signals could be used to test it. In the short
term, searches for disappearing tracks or fermiophobic scalars are the
most promising for probing part of the parameter space. Searches for
the third generation of squarks are also important but it is
challenging to apply their bounds to the present scenario where
squarks have multi-body decays~\cite{Alves:2013wra}. 

\section*{Acknowledgments}

I am grateful to the organizers of the Rencontres de Blois for the pleasant atmosphere. I would also like to thank
A. Delgado, G. Nardini and M. Quir\'os for the fruitful collaboration.

\end{document}